\documentclass[intlimits,twoside,a4paper]{article}

\usepackage[cp1251]{inputenc}
\usepackage[eqsecnum]{cmpj3}

\usepackage{bm}
\usepackage{footmisc}

\issue{2021}{24}{1}{13706}
\doinumber{10.5488/CMP.24.13706}
\title[The Hexagonal MAuGe (M= Lu, Sc) compounds]%
{An ab initio study of  structural, elastic and electronic properties of  hexagonal MAuGe (M= Lu, Sc) compounds%
\thanks{E-mail: mradjai@yahoo.com}}
\author[M. Radjai, N. Guechi, D. Maouche]{M. Radjai\refaddr{label1},
        N. Guechi\refaddr{label2,label3}, D. Maouche\refaddr{label4}}
\addresses{
\addr{label1} Laboratory of Physics of Experimental Techniques and Their Applications (LPTEAM), University of Medea, Algeria
\addr{label2} Laboratoire d'Etudes des Surfaces et Interfaces des Materiaux Solides, University Ferhat Abbas Setif 1, Algeria
\addr{label3} Faculty of Medcine, University Ferhat Abbas Setif 1, Algeria.
\addr{label4} Laboratory for Developing New Materials and Their Characterizations, University Ferhat Abbas Setif 1, Algeria 
}
\date{Received April 17, 2020, in final form October 9, 2020}
\begin{document}

\maketitle

\begin{abstract}
In this paper, we  performed a detailed theoretical study of structural, elastic and electronic properties of two germanides LuAuGe and ScAuGe by means of first-principles calculations using the pseudopotential plane-wave method within the generalized gradient approximation. The crystal lattice parameters and the internal coordinates are in good agreement with the existing experimental and theoretical reports, which proves the reliability of the applied theoretical method. The hydrostatic pressure effect on the structural parameters is shown. The monocrystalline elastic constants were calculated using the stress-strain technique. The calculated elastic constants of the  MAuGe  (M = Lu, Sc)  compounds meet the mechanical stability criteria for hexagonal crystals and these constants were used to analyze the elastic anisotropy of the MAuGe compounds through three different indices. Polycrystalline isotropic elastic moduli, namely bulk modulus, shear modulus, Young's modulus, Poisson's ratio, and the related properties are also estimated using Voigt-Reuss-Hill approximations. Finally, we  studied the electronic properties of the considered compounds by calculating their band structures, their densities of states and their electron density distributions.


%\textbf{PACS}: 71.15.Ap, 71.15.Mb, 71.15.Nc, 71.20.Nr, 65.40.Ba, 65.40.De,
%78.20.Ci

\end{abstract}%

\section{Introduction}

In the recent years,  ternary intermetallic compounds MAuGe (M denoting a rare earth element) have received an increased scientific attention due to their fascinating structural variety, exceptional physical properties, and wide range of applications. As a result, many scientific reports were published on their crystal structures and physical properties \cite{Pottgen96,Fornasini92,Tsetseris17,Schnelle97}. Ternary germanides, for example, provide a wider range of interest in magnetic susceptibility, electrical resistivity and
specific heat \cite{Schnelle97}. Note that the physical properties of the MAuGe ternary germanides
strongly depend on the nature of the rare earth element M.  According to the authors of the reference 
\cite{Pottgen96}, the compounds ScAuGe  and LuAuGe are diamagnetic materials and exhibit remarkable physical properties which make them interesting for possible technological applications.

P\"{o}ttgen R. et al. \cite{Pottgen96} recently reported the experimental preparation of the new germanides MAuGe (M~=~Lu, Sc) by melting alloys prepared from their atomic constituents in an arc furnace with the subsequent annealing at 1070~K. The crystal structures of MAuGe (M = Lu, Sc) compounds were determined by X-ray diffraction. They show crystal structures derived from the CaIn$_{2}$ structure-type by an ordered arrangement of Au and Ge atoms in the In positions. Note  that the MAuGe (M = Lu, Sc) have crystal structures similar to that of the LiGaGe compound \cite{Bockelnn70, Bockelnn74}. In addition, the MAuGe compounds crystallize in the  P$6_ {3} mc $ space group  (No.~186)  where the  Au  and  Ge  atoms form a three-dimensional $(3D)$-[AuGe] polyanions of elongated tetrahedra in MAuGe \cite{Pottgen96}. The frowning degree of the [AuGe] hexagonal greately depends on the size of the atom M = (Lu, Sc), so that the lattice constant $c$ of the hexagonal lattice increases systematically with the size of the atom  M  (M = Lu, Sc). Schnelle et al. \cite{Schnelle97} classify the MAuGe (M = Lu, Sc) as weak diamagnetics, and measurements of their electrical resistivities indicate a metallic behavior for both compounds.

To the best of the authors' knowledge, no theoretical or experimental study of the elastic properties of these compounds has been carried out up till now. Certainly, it is very important to get to know the elastic properties of a material  because they provide information on the stability and stiffness of the material against the externally applied stresses. Due to the close relationships of elastic properties with other fundamental physical properties, in the present work we performed detailed ab initio calculations of structural, elastic and electronic properties of MAuGe with  M = (Lu, Sc) under the pressure effect. Note that the measurements of elastic and structural parameters under pressure are generally difficult to determine experimentally. Therefore, to know the elastic constants and the evolution of the lattice parameters under the pressure effect is very important in many modern technologies \cite{Zhijiao11}. We hope that the reported results in this article will be useful for further studies or for possible technological applications of the MAuGe germanides.

\section{Computational details}

All quantum mechanics calculations were performed using the
pseudopotential plane wave (PP-PW) method in the framework of the
density functional theory (DFT) as implemented in the CASTEP code
(CambridgeSerial Total Energy Package) \cite{Clark05}. In order to calculate the
structural parameters and elastic moduli properties, the exchange
correlation energy is treated within the generalized gradient approximation 
GGA-PBEsol as parameterized by Perdew et al. \cite{Perdew08}. For all electronic total energy calculations, the valence electrons of the Lu,  Sc, Au and Ge pseudo-atoms are described by the Vanderbilt-ultra-soft pseudopotential \cite{Vanderbilt92}. The Lu $4f^{14}$ $5p^{6}$ $5d^{1}$ $6s^{2}$, Au $5d^{10}$ $6s^{1}$, Ge $4s^{2}$ $4p^{2}$ and Sc $3s^{2}$ $3p^{6}$ $3d^{1}$ $4s^{2}$ orbitals are explicitly treated as valence states.
The plane-wave basis set was defined by a plane-wave cut-off energy of 400~eV, and the Brillouin zone (BZ) integration was performed over the $5\times 5\times 4$ grid size using
Monkhorst-Pack scheme \cite{Monkhorst76} for hexagonal structure. The
plane-wave basis set and the grid size were chosen after a convergence test in
order to ensure a sufficiently accurate converged total energy, thus optimizinng the geometry, computing the elastic constants and the electronic structures of MAuGe.
The Broyden--Fletcher--Goldfarb--Shanno (BFGS) minimization technique~\cite{Fischer92} was used to determine the structural parameters of the equilibrium geometries because the (BFGS) method provides a way to find the lowest energy of the considred cristalline structure. This set of parameters was carried out with the following convergence criteria: a maximum of an ionic Hellmann--Feynman force within  $5\times10^{-2}$ eV/\AA, a maximum stress of $10^{-1}$~GPa, a maximum displacement of $2\times10^{-3}$\AA~and a self-consistent convergence of the total energy of $2\times10^{-5}$ eV/atom. The well-known stress-strain approach~\cite{Guechi14,Milman01} was used to determine the elastic constants by applying a set of a difined homogeneous deformation with a finite value and by calculating the resulting stresses in the optimized and relaxed structures. The convergences criteria during the relaxation stage of the internal coordinates were chosen as follows: a total energy less than $4\times10^{-6}$ eV/atom, a converged forces within $10^{-2}$ eV/\AA~and a maximum ionic displacement of $4\times10^{-4}$
\AA.

\section{Results and discussion}

\subsubsection{Structural properties}
\begin{figure}[!t] %\centering%
	\begin{center}
       \includegraphics[scale=0.92]{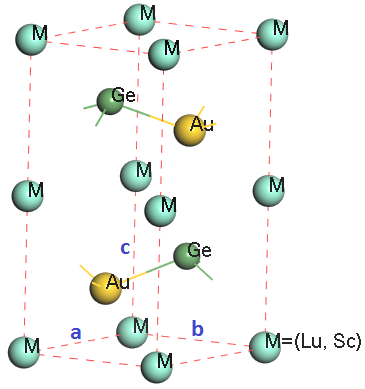}
    \end{center}
	\caption{(Colour online) The unit-cell crystalline structure of the MAuGe compounds (M= Lu, Sc).}
	\label{fig1}
\end{figure}
The ternary compounds LuAuGe and ScAuGe  crystallize in a hexagonal structure type 
with the space group $P6_{3}/mc$ (No.~186) \cite{Pottgen96}. The conventional cell of the MAuGe (M = Lu, Sc) germanides contains two chemical formulae $(Z=2)$. Therefore, it contains $6$ atoms per unit cell as show in { figure~\ref{fig1}}. The M, Au and Ge atoms occupy the following Wyckoff  positions:  M=(Lu, Sc):
$2a$ ($0$, $0$, $z_{\text M}$), Au: $2b$ ($1/3$, $2/3$, $z_{\text{Au}}$) and Ge: $2b$
(1/3, $2/3$, $z_{\text{Ge}})$, where $z_{\text M}$, $z_{\text{Au}}$ and $z_{\text{Ge}}$ are the internal $z$-coordinates of M=(Lu, Sc), Au and Ge atoms, respectively. Thus, the MAuGe unit cell is characterized by five structural parameters not fixed by symmetry: two lattice constants (%
$a$ and $c$) and three internal coordinates ($z_{\text M}$, $z_{\text{Au}}$ and $z_{\text{Ge}}$).

%\scriptsize
\begin{table}[!t]
\caption{Calculated equilibrium crystal lattice constants (a and c, in
\AA) and volume (V, in \AA$^3$) for the MAuGe(M=Lu, Ge) compounds compared with the available experimental and theoretical data.}
\vspace{3mm}
\centering
\begin{tabular}[b]{|c|c|c|c|c|c|c|}
\hline
{\scriptsize Structural} & \multicolumn{3}{|c}{LuAuGe} &
\multicolumn{3}{|c|}{{ Sc}AuGe} \\ \cline{2-7}
{\scriptsize Parameter} & \multicolumn{1}{|c|}{\scriptsize Present work} &
\multicolumn{1}{|c|}{{\scriptsize Expt.\cite{Pottgen96}}} &{\scriptsize Other} &
{\scriptsize Present work} & {\scriptsize Expt.\cite{Pottgen96}} &{\scriptsize Other}
\\ \hline
${\scriptsize a}$ & 4.418 & 4.377 & 4.337\cite{Schnelle97} & 4.332 & 4.382 & 4.377\cite{Tsetseris14}, 4.308\cite{Schnelle97} \\ \hline
${\scriptsize c}$ & 7.032 & 7.113 & 7.113\cite{Schnelle97} & 6.796 & 6.845 & 7.083\cite{Tsetseris14}, 6.845\cite{Schnelle97} \\ \hline
$\frac{c}{a}$ & 1.591 & 1.625 & 1.625\cite%
{Schnelle97} & 1.568 & 1.620 & 1.618 \cite{Tsetseris14}, 1.588 \cite{Schnelle97} \\ \hline
${\scriptsize V}$ & 118.8 & 118.1 & 118.1\cite{Schnelle97} & 110.4 & 110.0 & - \\\hline
\end{tabular}
\label{tab1}
\end{table}

As the first step of our calculations, we used the experimental structural parameters in order to calculate the optimized lattice constants ($a$ and $c$) and the internal atomic $z$-coordinates at zero pressure. The calculated structural parameters of MAuGe, including the equilibrium lattice
constants, $a_{0}$ and $c_{0}$, and the internal coordinates, $z_{M_{0}}$, $z_{\text{Au}_{0}}$ and $z_{\text{Ge}_{0}}$, using the PP-PW method within the GGA approximation are shown in table~\ref{tab1} and table~\ref{tab2} in comparison  with experimental \cite{Pottgen96} and theoretical data \cite{Tsetseris14,Schnelle97}.

\begin{table}[!t]
\caption{Calculated internal atomic coordinates for the hexagonal compounds MAuGe
(M = Lu, Ge) in comparison with experiment.}
\vspace{3mm}
\centering
\begin{tabular}[b]{cc|c|c|c|c|c|c|}
\cline{3-8}
&  & \multicolumn{3}{|c|}{LuAuGe} & \multicolumn{3}{|c|}{{ Sc%
}AuGe} \\ \cline{3-8}\cline{7-7}
&  & {\normalsize x} & {\normalsize y} & {\normalsize z} & {\normalsize x} & 
{\normalsize y} & {\normalsize z} \\ \hline
\multicolumn{1}{|c}{ Lu} & \multicolumn{1}{|c|}{ %
Present} & {\normalsize 0} & {\normalsize 0} & {\normalsize 0.9940} & 
{\normalsize -} & {\normalsize -} & {\normalsize -} \\ \cline{2-8}\cline{7-7}
\multicolumn{1}{|c}{\normalsize (2e)} & \multicolumn{1}{|c|}{{\normalsize %
Expt\cite{Pottgen96}}} & {\normalsize 0} & {\normalsize 0} & {\normalsize %
0.9941} & {\normalsize -} & {\normalsize -} & {\normalsize -} \\ \hline
\multicolumn{1}{|c}{ Sc} & \multicolumn{1}{|c|}{\normalsize %
Present} & {\normalsize -} & {\normalsize -} & {\normalsize -} & 
{\normalsize 0} & {\normalsize 0} & {\normalsize 0.00124} \\ 
\cline{2-8}\cline{7-7}
\multicolumn{1}{|c}{\normalsize (2a)} & \multicolumn{1}{|c|}{{\normalsize %
Expt\cite{Pottgen96}}} & {\normalsize -} & {\normalsize -} & {\normalsize -}
& {\normalsize 0} & {\normalsize 0} & {\normalsize 0.0012} \\ \hline
\multicolumn{1}{|c}{ Au} & \multicolumn{1}{|c|}%
{\normalsize Present} & {\normalsize 0.333} & {\normalsize 0.666} & 
{\normalsize 0.6999} & {\normalsize 0.333} & {\normalsize 0.666} & 
{\normalsize 0.6999} \\ \cline{2-8}\cline{7-7}
\multicolumn{1}{|c}{\normalsize (2b)} & \multicolumn{1}{|c|}{{\normalsize %
Expt\cite{Pottgen96}}} & {\normalsize 0.333} & {\normalsize 0.666} & 
{\normalsize 0.7000} & {\normalsize 0.333} & {\normalsize 0.666} & 
{\normalsize 0.7000} \\ \hline
\multicolumn{1}{|c}{ Ge} & \multicolumn{1}{|c|}{\normalsize %
Present} & {\normalsize 0.333} & {\normalsize 0.666} & {\normalsize 0.2887}
& {\normalsize 0.333} & {\normalsize 0.666} & {\normalsize 0.298} \\ 
\cline{2-8}\cline{7-7}
\multicolumn{1}{|c}{\normalsize (2b)} & \multicolumn{1}{|c|}{{\normalsize %
Expt\cite{Pottgen96}}} & {\normalsize 0.333} & {\normalsize 0.666} & 
{\normalsize 0.2886} & {\normalsize 0.333} & {\normalsize 0.666} & 
{\normalsize 0.298} \\ \hline
\end{tabular}
\label{tab2}
\end{table}%
%\end{center}

As can be seen from table~\ref{tab1} and table~\ref{tab2}, our calculated equilibrium structural parameters ($a_{0}$, $c_{0}$, $z_{\text M}$, $z_{\text{Au}}$ and $z_{G_{e}}$) are in very
good agreement with the existing experimental and theoretical data. The calculated values of the five optimized structural parameters of LuAuGe (ScAuGe) deviate from the measured ones by less than $0.93\%,1.13\%,0.01\%,0.01\%$ and $0.01\%$ ($0.93\%,1.13\%,0.01\%,0.01%
\% $ and $0.01\%$), respectively. Moreover, the calculated and measured internal atomic  coordinates of M, Au and Ge atoms of the cell unit are in very good agreement with the experimental and theoretical values \cite{Pottgen96,Tsetseris14,Schnelle97}. This excellent matching is an indication
of the capability of this chosen first-principles method to provide confidence for the calculations of the elastic and electronic properties of the titled compounds.

\begin{figure}[!b] %\centering%
	\begin{center}
       \includegraphics[scale=0.45]{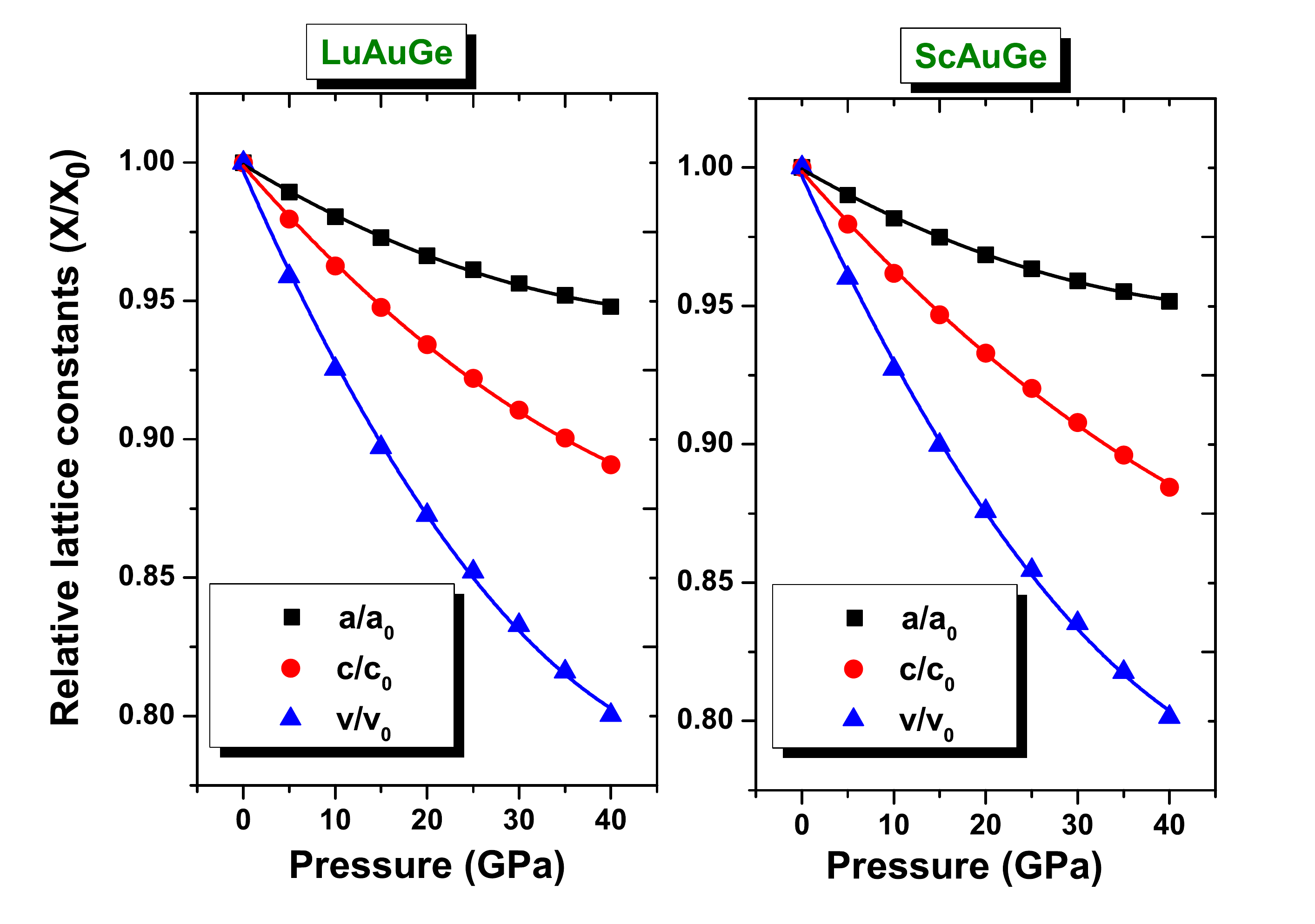}
    \end{center}
	\caption{(Colour online) Pressure dependence of the relative variations of the lattice
constants ($a$ and $c$) and volume $V$ for the LuAuGe and ScAuGe
compounds. The index 0 indicates the structural parameter value at zero pressure.}
	\label{fig2}
\end{figure}

To evaluate the hydrostatic pressure effect on the structural
properties ($a$, $c$ and $V$) of MAuGe (M = Lu, Sc), in figure~\ref{fig2} we illustrated in their relative changes $X/X_{0}$ under pressure, where $X$ represents $a$, $c$ or $V$, and $X_{0}$ refers to the structural parameter value at zero pressure. The evolution of the relative changes of each parameter in the considered pressure range was well fitted to a second order polynomial~\cite{Guechi14}:
\begin{equation}
\frac{X\left( P\right) }{X_{0}}=1+\beta _{X}P+K P^{2},
\end{equation}
where $X=a, c$ or $V$ , $\beta _{X}=-\frac{1}{X}\frac{\rd x}{\rd p}$ and $K$ is a constant.

The expressions describing the relative changes of structural parameters $a$, $c$ and $V$ are as follows:

\begin{equation}
\text{LuAuGe}\left\{
\begin{array}{l}
\frac{a}{a_{0}}=1-2.02\times10^{-3}P+1.87\times10^{-5}\text{ }P^{2}\\
\frac{c}{c_{0}}=1-3.80\times10^{-3}P+2.80\times10^{-5}P^{2} \\
\frac{V}{V_{0}}=1-7.60\times10^{-3}P+6.83\times10^{-5}P^{2}%
\end{array}%
\right.  \label{(6)}
\end{equation}

\begin{equation}
\text{ScAuGe}\left\{
\begin{array}{l}
\frac{a}{a_{0}}=1-1.90\times10^{-3}P+1.80\times10^{-5}P^{2} \\
\frac{c}{c_{0}}=1-3.75\times10^{-3}P+2.31\times10^{-5}P^{2} \\
\frac{V}{V_{0}}=1-7.34\times10^{-3}P+6.25\times10^{-5}P^{2}%
\end{array}%
\right.  \label{7}
\end{equation}

It can easily be seen from figure~\ref{fig2} that the relative variation of  structural parameters decrease when the pressure goes from $0$~GPa to $40$~GPa. The estimated
values of the linear compressibilities of the lattice parameters $a$ and $c$
are $\beta _{a}=$ $-2.02\times10^{-3}GPa^{-1}$ and $\beta _{c}$ $%
=-3.8\times10^{-3}$GPa$^{-1}$ for LuAuGe and $\beta _{a}=-1.9\times10^{-3}$GPa$^{-1}$ and
$\beta _{c}$ $=-3.75\times10^{-3}$GPa$^{-1}$ for ScAuGe. We also see that for the two materials, $\beta _{c}$ is less than $\beta _{a}$ which implies that the 
$c/c_{0}$ ratio decreases faster than the $a/a_{0}$ ratio. Therefore, the MAuGe compounds are relatively more compressible along the $c$-axis
than along the $a$-axis. It hould also be noted that the clearly different values of $\beta _{a}$ and $\beta _{c}$ reveal a notable compression anisotropy.

The linear ($\beta _{a}$ and $\beta _{c}$) and volumic $\beta _{V}$ compressibilities  obtained from the lattice parameters ($a$ and $c$) and volume $V$, respectively, were used to estimate the bulk modulus $B$ as follows \cite{Milman01}:

\begin{equation}
B=\frac{1}{2\beta _{a}+\beta _{c}},  \label{3.4}
\end{equation}

\begin{equation}
B=\frac{1}{\beta_{V}}.  \label{3.5}
\end{equation}

\begin{figure}[!t] %\centering%
	\begin{center}
       \includegraphics[scale=0.5]{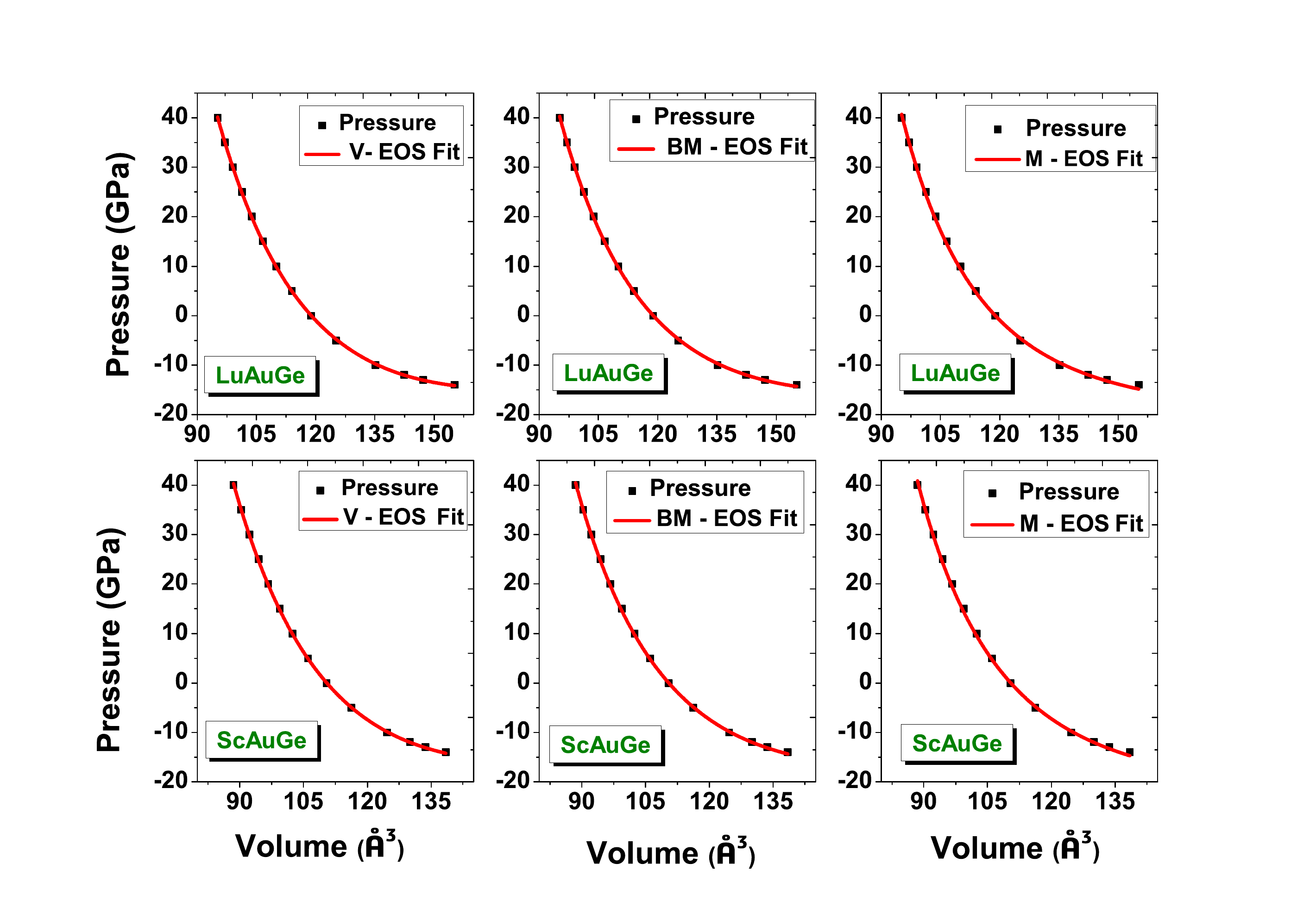}
    \end{center}
	\caption{(Colour online) Calculated pressure vs. volume $P(V)$ for the hexagonal compounds
LuAuGe and ScAuGe. The solid lines are the fits of the obtained data to the Vinet (V-EOS), Birch Murnaghan (BM-EOS) and Murnaghan equations of states (M-EOS).}
\label{fig3}
\end{figure}

\begin{figure}[!t] %\centering%
	\begin{center}
      \includegraphics[scale=0.4]{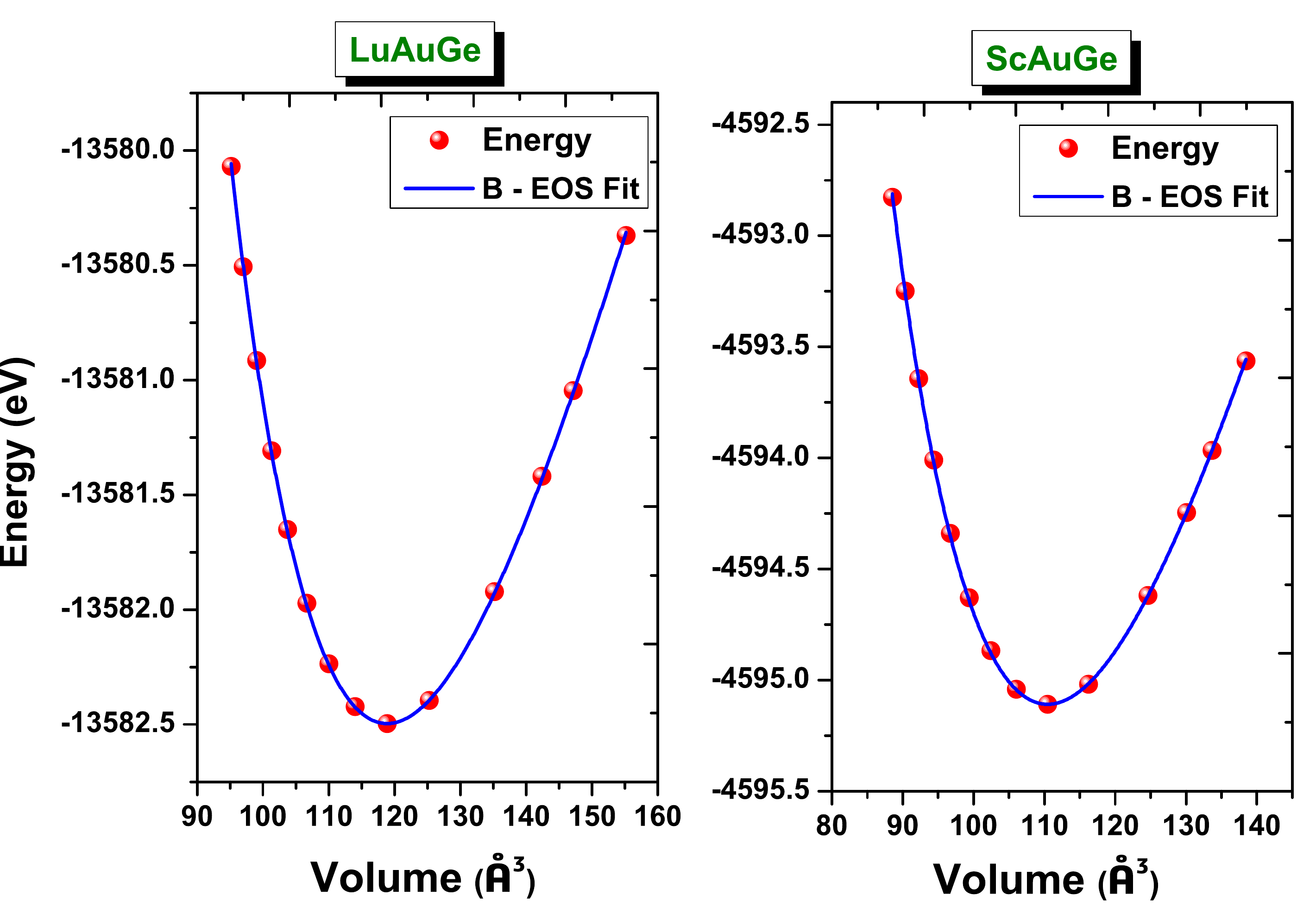}
    \end{center}
	\caption{(Colour online) Calculated total energy vs. volume $E(V)$  for the hexagonal compouds
LuAuGe and ScAuGe. The solid lines are the fits of the obtained data to the
Birch equation of states (B-EOS).}
\label{fig4}
\end{figure}

The calculted values of the bulk modulus $B$ using the equations (\ref{3.4}) and (\ref{3.5}) are listed in table~\ref{tab3}. The bulk modulus can also be extracted from the fit of the data Energy-Volume $ (E-V) $ (Pressure-Volume $ (P-V) $) by the equations of states EOS  of solid materials which describe the variation of the energy (pressure) as a function of volume. The calculated data $ (E-V) $ and $ (P-V) $ are very well adjusted to the following $ (EOS) $: Birch EOS  \cite{Birch47}, Birch-Murnaghan EOS  \cite{Birch78, Ambrosch-Draxl06}, Vinet EOS  \cite{Vinet89, Fu83} and Murnaghan EOS  \cite{Murnaghan44} to determine the bulk modulus $ B $ and its derivative with respect to the pressure $B^{\prime}$. The obtained results are visualized in figure~\ref{fig3} and  figure~\ref{fig4} and tabulated in {\normalsize table 3}. We can obseve that there is a good agreement between different procedures used to evaluated the bulk modulus $B$.  Therefore, these results constitute a good proof of the reliability for our calculations. 

\begin{table}[!b]%
%EndExpansion
\caption{Calculated bulk modulus ($B$, in GPa) and its pressure derivative $B'$.}
\centering
\vspace{3mm}
\begin{tabular}[b]{c|c|c|c|c|c|c|}
\cline{2-7}
& \multicolumn{3}{|c|}{LuAuGe} & \multicolumn{3}{|c|}{ScAuGe} \\ \hline
\multicolumn{1}{|c|}{ $B$} & {\normalsize 105.38}$^{a}$ & {\normalsize %
100.28}$^{b}$ & {\normalsize 103.81}$^{c}$ & {\normalsize 109.36}$^{a}$ &
{\normalsize 105.37}$^{b}$ & {\normalsize 108.03}$^{c}$ \\ \cline{2-7}
\multicolumn{1}{|c|}{} & {\normalsize 105.45}$^{d}$ & {\normalsize 113.12}$%
^{e}$ & {\normalsize 114.5}$^{f}$ & {\normalsize 110.45}$^{d}$ &
{\normalsize 117.23}$^{e}$ & {\normalsize 118.62}$^{f}$ \\ \hline
\multicolumn{1}{|c|}{ $B'$} & {\normalsize 4.96}$^{a}$ & {\normalsize %
4.94}$^{b}$ & {\normalsize 4.93}$^{c}$ & {\normalsize 4.75}$^{a}$ &
{\normalsize 4.66}$^{b}$ & {\normalsize 4.37}$^{c}$ \\ \cline{2-7}
\multicolumn{1}{|c|}{} & {\normalsize 5.02}$^{d}$ & {\normalsize -} &
{\normalsize -} & {\normalsize 4.91}$^{d}$ & {\normalsize -} & {\normalsize -%
} \\ \hline
\end{tabular}%
\label{tab3}
\end{table}%

$^{a}$ Calculated from Vinet EOS \cite{Fu83}

$^{b}$ Calculated from Murnaghan EOS \cite{Birch78}

$^{c}$ Calculated from Birch-Murnaghan EOS \cite{Ambrosch-Draxl06}

$^{d}$ Calculated from Birch EOS \cite{Birch47}

$^{e}$ Calculated from linear compressibilities $ \beta _{a} $ and $ \beta _{c} $: $B=1/(2\beta _{a}+\beta _{c})$ 

$^{f}$ Calculated from the volumic compressibility $ \beta _{V} $: $B=1/\beta _{V}$.

\begin{figure}[!t] %\centering%
	\begin{center}
      \includegraphics[scale=0.5]{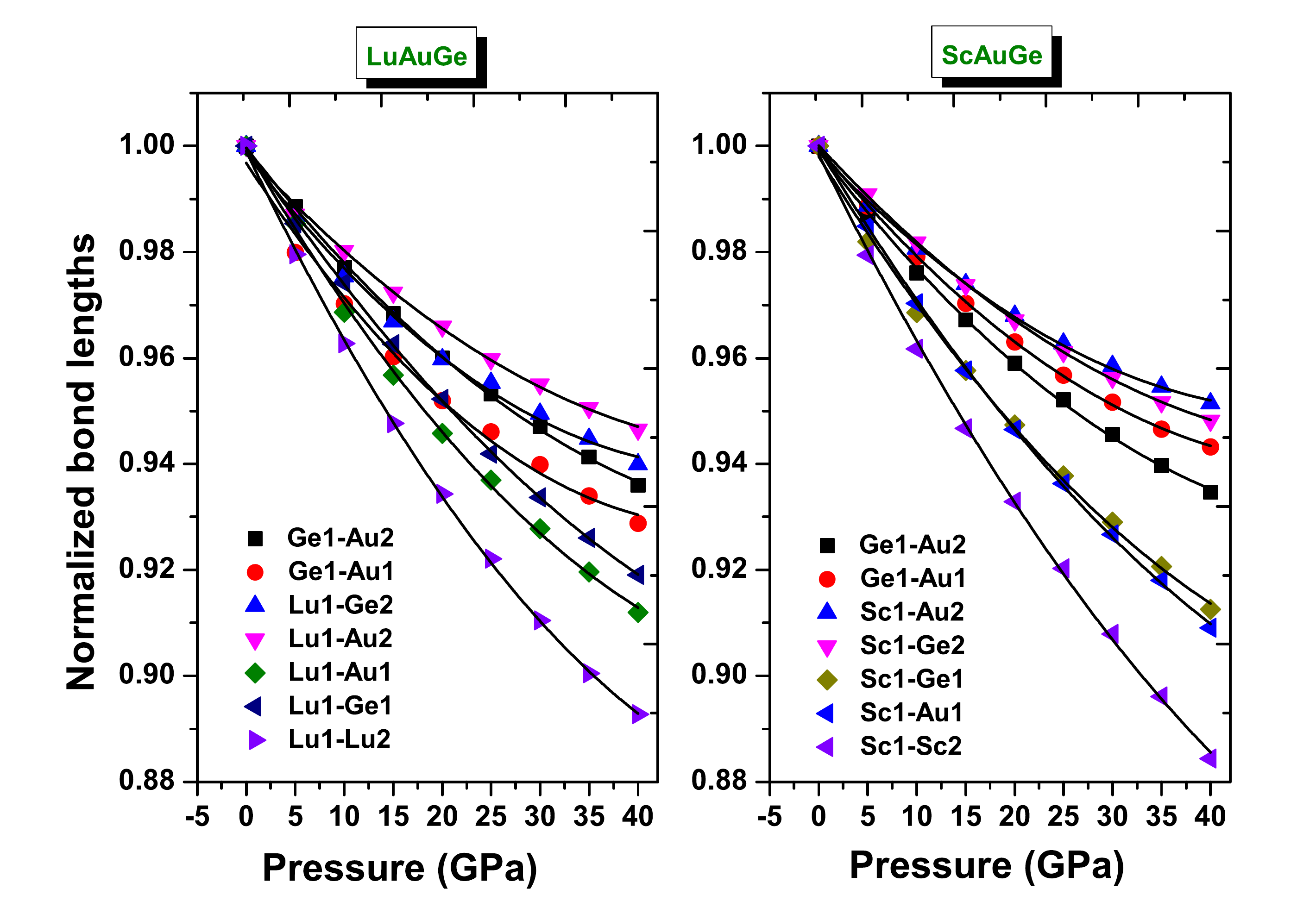}
    \end{center}
	\caption{(Colour online) Pressure-dependent variations of the relative bond-lengths in LuAuGe and ScAuGe materials.}
	\label{fig5}
\end{figure}

In order to fully characterize the pressure dependence of the structural
parameters, we studied the pressure effect on the bond lengths between the following first atomic neighbours: Ge$_{1}$, Ge$_{2}$, Au$_{1}$, Au$_{2}$, Lu$_{1}$, Lu$_{2}$, Sc$_{1}$ and Sc$_{2}$. The pressure effect on the normalized bond lengths ${L}/{L_{0}}$ is plotted in  figure~\ref{fig5}. ${L}$ stands for the bond-length at pressure $P$, and $L_{0}$ is its corresponding value at zero pressure. 

From figure~\ref{fig5}, we observe that a different bond-length as a fuction of pressure in MAuGe compounds perfectly follows the second order polynomial as shown in  equations~(\ref{eq3.6}) and (\ref{eq3.7}). The calculated bond lengths at zero pressure are listed in {\normalsize table 4} with the available experimental data \cite{Pottgen96}. From the reported results in { table~\ref{tab4}}, it can be seen that the Lu$_{1}$-Lu$_{2}$ and Sc$_{1}$-Sc$_{2}$ bonds are more compressible than the other bonds, while the Ge$_{1}$-Au$_{2}$ bond in the two compounds is the least compressible. Note that our results are in good agreement with the reported experimental findings~\cite{Pottgen96}.

\begin{table}[!t]%
\caption{The calculated first order interatomic distances (in \AA) in MAuGe (M=Lu, Sc).}
\centering
\vspace{3mm}
\begin{tabular}[b]{|c|c|c|c|c|c|}
\hline
& \multicolumn{2}{|c}{LuAuGe} & \multicolumn{3}{|c|}{ScAuGe%
} \\ \cline{2-6}
& {\normalsize Present work} & {\normalsize Expt\cite{Pottgen96}} &  &
{\normalsize Present work} & {\normalsize Expt\cite{Pottgen96}} \\ \hline
Ge$_{1}$-Au$_{2}$ & 2.626 & 2.605 & Ge$_{1}$-Au$_{2}$ & 2.588 & 2.576 \\ \hline
Ge$_{1}$-Au$_{1}$ & 2.892 & 2.927 & Ge$_{1}$-Au$_{1}$ & 2.731 & 2.752 \\ \hline
 Lu$_{1}$-Ge$_{{2}}$ &
2.931 & 2.920 &  Sc$_{1}${-Au}$_{{2}}$ & 2.842 & 2.835 \\ \hline
 Lu$_{1}${-Au}$_{2}$ &
2.933 & 2.921 &  Sc$_{1}$-Ge$_{2}$ & 2.857 & 2.850 \\ \hline
 Lu$_{1}$-Au$_{1}$ &
3.283 & 3.281 &  Sc$_{1}${-Ge}$_{1}$ & 3.213 & 3.212 \\ \hline
 Lu$_{1}$-Ge$_{1}$ &
3.286 & 3.283 &  Sc$_{1}${-Au}$_{1}$ & 3.232 & 3.231 \\ \hline
 Lu$_{1}$-Lu$_{2}$ &
3.516 & 3.557 & Sc$_{1}${-Sc}$_{2}$ & 3.398 & 3.423 \\ \hline
\end{tabular}%
\label{tab4}
\end{table}%
\newpage
\begin{equation}
\text{LuAuGe}\left\{
\begin{array}{l}
\left( \frac{L}{L_{0}}\right)
_{\text{Ge}_{1}-\text{Au}_{2}}=1-2.37\times10^{-3}P+1.98\times10^{-5}P^{2} \\
\left( \frac{L}{L_{0}}\right)
_{\text{Ge}_{1}-\text{Au}_{1}}=1-2.83\times10^{-3}P+2.92\times10^{-5}P^{2} \\
\left( \frac{L}{L_{0}}\right)
_{\text{Lu}_{1}-\text{Ge}_{2}}=1-2.41\times10^{-3}P+2.45\times10^{-5}P^{2} \\
\left( \frac{L}{L_{0}}\right)
_{\text{Lu}_{1}-\text{Au}_{2}}=1-2.01\times10^{-3}P+1.80\times10^{-5}P^{2} \\
\left( \frac{L}{L_{0}}\right)
_{\text{Lu}_{1}-\text{Au}_{1}}=1-3.19\times10^{-3}P+2.55\times10^{-5}P^{2} \\
\left( \frac{L}{L_{0}}\right)
_{\text{Lu}_{1}-\text{Ge}_{1}}=1-2.76\times10^{-3}P+1.86\times10^{-5}P^{2} \\
\left( \frac{L}{L_{0}}\right)
_{\text{Lu}_{1}-\text{Lu}_{2}}=1-3.87\times10^{-3}P+3.03\times10^{-5}P^{2}%
\end{array}%
\right.
\label{eq3.6}
\end{equation}

\begin{equation}
\text{ScAuGe}\left\{
\begin{array}{l}
\left( \frac{L}{L_{0}}\right)
_{\text{Ge}_{1}-\text{Au}_{2}}=1-2.42\times10^{-3}P+2.06\times10^{-5}P^{2} \\
\left( \frac{L}{L_{0}}\right)
_{\text{Ge}_{1}-\text{Au}_{1}}=1-2.25\times10^{-3}P+2.11\times10^{-5}P^{2} \\
\left( \frac{L}{L_{0}}\right)
_{\text{Sc}_{1}-\text{Ge}_{2}}=1-2.00\times10^{-3}P+1.76\times10^{-5}P^{2} \\
\left( \frac{L}{L_{0}}\right)
_{\text{Sc}_{1}-\text{Au}_{2}}=1-1.95\times10^{-3}P+1.93\times10^{-5}P^{2} \\
\left( \frac{L}{L_{0}}\right)
_{\text{Sc}_{1}-\text{Au}_{1}}=1-3.06\times10^{-3}P+2.04\times10^{-5}P^{2} \\
\left( \frac{L}{L_{0}}\right)
_{\text{Sc}_{1}-\text{Ge}_{1}}=1-3.01\times10^{-3}P+2.24\times10^{-5}P^{2} \\
\left( \frac{L}{L_{0}}\right)
_{\text{Sc}_{1}-\text{Sc}_{2}}=1-3.75\times10^{-3}P+2.32\times10^{-5}P^{2}.%
\end{array}%
\right.
\label{eq3.7}
\end{equation}

\subsection{Elastic properties}

\subsubsection{Single-crystal elastic constants}

The elastic constants $C_{ij}s$ are important physical parameters for solid crystalline materials. In particular, they provide an information on the response of the material when an external mechanical stress is applied and regarding the nature of the forces acting in solid materials \cite{Westbrook00}. A crystalline solid material of a hexagonal symmetry is described by five independent elastic constants, namely $C_{11}$, $C_{33}$, $C_{44}$, $C_{12}$ and $C_{13}$. The calculated
numerical values of the five elastic constants at zero pressure are listed in table~\ref{tab5} for the LuAuGe and ScAuGe compounds. Note that there are no experimental or theoretical
results available in the literature for the elastic constants $C_{ij}s$ of the MAuGe compounds (M = Lu, Sc) to be compared with our results. The present work
is the first attempt to calculate the elastic constants $C_{ij}s$ of the
titled compounds.

\begin{table}[!b]%
\caption{Calculated elastic constant  $C_{ijs}$ (in GPa) for the MAuGe (M =
Lu, Sc) compounds.}
\centering
\vspace{3mm}
\begin{tabular}{|c|c|c|c|c|c|}
\hline
Compounds & $C_{11}$ & $C_{33}$ & $C_{44}$ & $C_{12}$ & $C_{13}$ \\ \hline
 LuAuGe & 203.92 & 160.56 & 57.62 & 94.46 & 65.34 \\ \hline
ScAuGe & 233.70 & 180.59 & 58.68 & 87.53 & 61.26 \\ \hline
\end{tabular}%
\label{tab5}
\end{table}%

\begin{figure}[!h] %\centering%
	\begin{center}
       \includegraphics[scale=0.5]{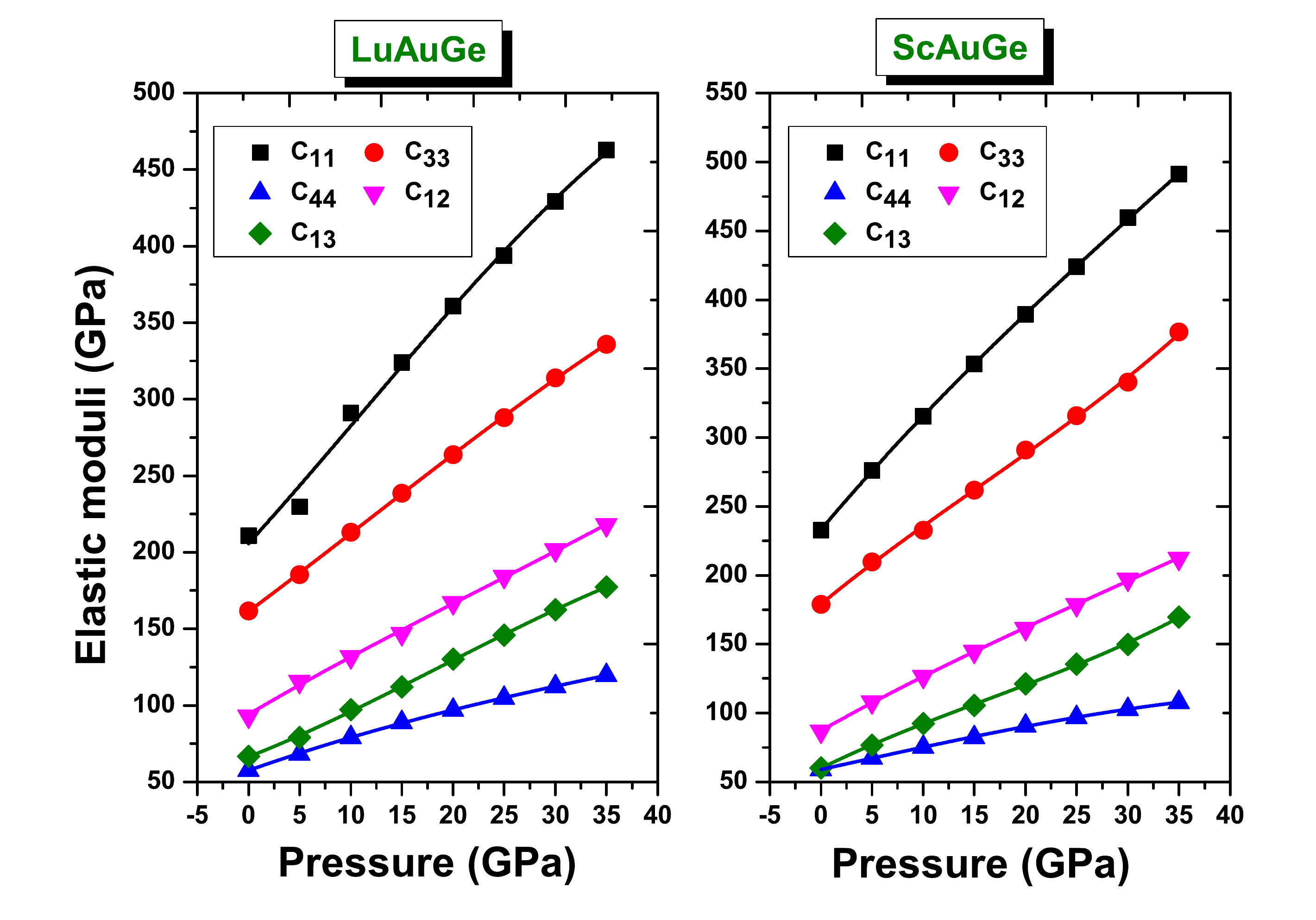}
    \end{center}
	\caption{(Colour online) Calculated pressure dependence of the independent elastic constants
C$_{ij}s$ for MAuGe (M=Lu, Sc). The symbols indicate the
calculated values. The lines represent the linear fitting curves.}
\label{fig6}
\end{figure}

From the obtained results, we can draw the following conclusions: (i)  we observe that  $C_{11}$ is greater than  $C_{44}$, $C_{12}$ and $C_{13}$, which shows that
the considered systems are more resistant to unidirectional compressions than
to shear strains; (ii) knowing that $ C_ {11} $ and $ C_ {33} $ reflect the
uniaxial stiffness along the $ a $ and $ c $ axes, respectively, the obtained $C_{11}$ value is greater than that of $C_{33}$  for the two compounds, indicating that MAuGe (M = Lu, Sc) are
relatively stiffer materials along the $a$ axis than along $c$ axis. This result agrees very well with the results already obtained by adjusting the relative changes of the lattice constants $a$ and $c$ as a function of pressure illustrated  in figure~\ref{fig2}; (iii) figure~\ref{fig6} shows the pressure dependence of the five independent elastic constants of the MAuGe compounds
(M = Lu, Sc) for pressures up to 40~GPa. It can be clearly seen  in figure~\ref{fig6} that the elastic constants $C_{ij}$ increases monotonously with the increasing pressure, and the fit results are given in equations (\ref{3.9}) and (\ref{3.10}). (iv) The  mechanical stability of MAuGe compounds (M = Lu, Sc) is verified because the $C_{ijs}$ calculated at zero pressure satisfy the following mechanical stability restrictions \cite{Wu07}:

\begin{equation}
\begin{array}{l}
C_{11}> 0,\quad\text{ }C_{11}-C_{12}> 0,\quad\text{ }C_{44}> 0,\quad\text{
}\left( C_{11}+C_{12}\right) C_{33}-2C_{13}^{2}> 0.%
\end{array}
\label{3.8}
\end{equation}

Thus, we can assert that the hexagonal MAuGe (M = Lu, Sc) is in a
mechanically stable state for pressure range 0--40~GPa.

The fit results are given by the following expressions for both
compounds LuAuGe and ScAuGe, respectively:

\begin{equation}
\text{LuAuGe}\left\{
\begin{array}{l}
C_{11}=203.92+8.27P-2.5\times10^{-2}P^{2} \\
C_{33}=160.56+5.31P-8.09\times10^{-3}P^{2} \\
C_{44}=57.63+2.24P-1.39\times10^{-2}P^{2} \\
C_{12}=94.46+3.72P-5.59\times10^{-3}P^{2} \\
C_{13}=65.34+3.12P-2.91\times10^{-3}P^{2}%
\end{array}%
\right.  \label{3.9}
\end{equation}

\begin{equation}
\text{ScAuGe}\left\{
\begin{array}{l}
C_{11}=233.70+8.42P-3.07\times10^{-2}P^{2} \\
C_{33}=180.59+5.27P-6.77\times10^{-3}P^{2} \\
C_{44}=58.68+1.77P-1.04\times10^{-2}P^{2} \\
C_{12}=87.53+3.95P-1.12\times10^{-2}P^{2} \\
C_{13}=61.26+2.90P-4.11\times10^{-3}P^{2}.%
\end{array}%
\right.
\label{3.10}
\end{equation}

Acoustic wave velocities in a material can be obtained from the Christoffel
equation \cite{Landau80}. The sound wave velocities propagating in the $%
[100] $, $[001]$ and $[120]$ directions in a hexagonal structure can be
calculated using the following relations:

\begin{equation}
\left\{
\begin{array}{c}
v_{L}^{[100]}=v_{L}^{[120]}=\sqrt{\frac{C_{11}}{\rho }},\text{ \ }%
v_{T_{1}}^{[100]}=v_{T_{1}}^{[120]}=\sqrt{\frac{\left( C_{11}-C_{12}\right)
}{2\rho }} \\
v_{T_{2}}^{[100]}=v_{T_{2}}^{[120]}=\sqrt{\frac{C_{44}}{\rho }} \\
v_{L}^{[001]}=\sqrt{\frac{C_{33}}{\rho }},\text{ \ \ \ \ }%
v_{T_{1}}^{[001]}=v_{T_{2}}^{[001]}=\sqrt{\frac{C_{44}}{\rho }},%
\end{array}%
\right.
\label{3.11}
\end{equation}
where $\rho $ is the mass density, $T$ and $L$ stand for transverse and
longitudinal polarizations, respectively. The calculated sound velocities at zero pressure extracted along $[100]$, $[120]$ and $[001]$ directions for MAuGe (M = Lu, Sc) are listed in table~\ref{tab6}

\begin{table}[!t]%
%EndExpansion
\caption{Acoustic wave velocities (in m/s) along different propagation
directions for MAuGe (M = Lu, Sc).}
\vspace{3mm}
\centering
\begin{tabular}[b]{|c|c|c|c|c|c|c|}
\hline
\text{System} &\strut $v_{L}^{[100]}=v_{L}^{[120]}$ &\strut $%
v_{T_{1}}^{[100]}=v_{T_{1}}^{[120]}$ &\strut $v_{T_{2}}^{[100]}=v_{T_{2}}^{[120]}$
& \strut$v_{L}^{[001]}$ & $v_{T_{1}}^{[001]}$ &\strut $v_{T_{2}}^{[001]}$ \\ \hline\hline
\text{LuAuGe} & 4104.53 & 2170.29 & 2146.13 & 3594.26 & 2146.13 & 2146.13 \\
\hline
\text{ScAuGe} & 4951.50 & 2772.46 & 2489.87 & 4342.29 & 2489.87 & 2489.87 \\
\hline
\end{tabular}%
\label{tab6}
\end{table}
\begin{figure}[!b] %\centering%
	\begin{center}
       \includegraphics[scale=0.4]{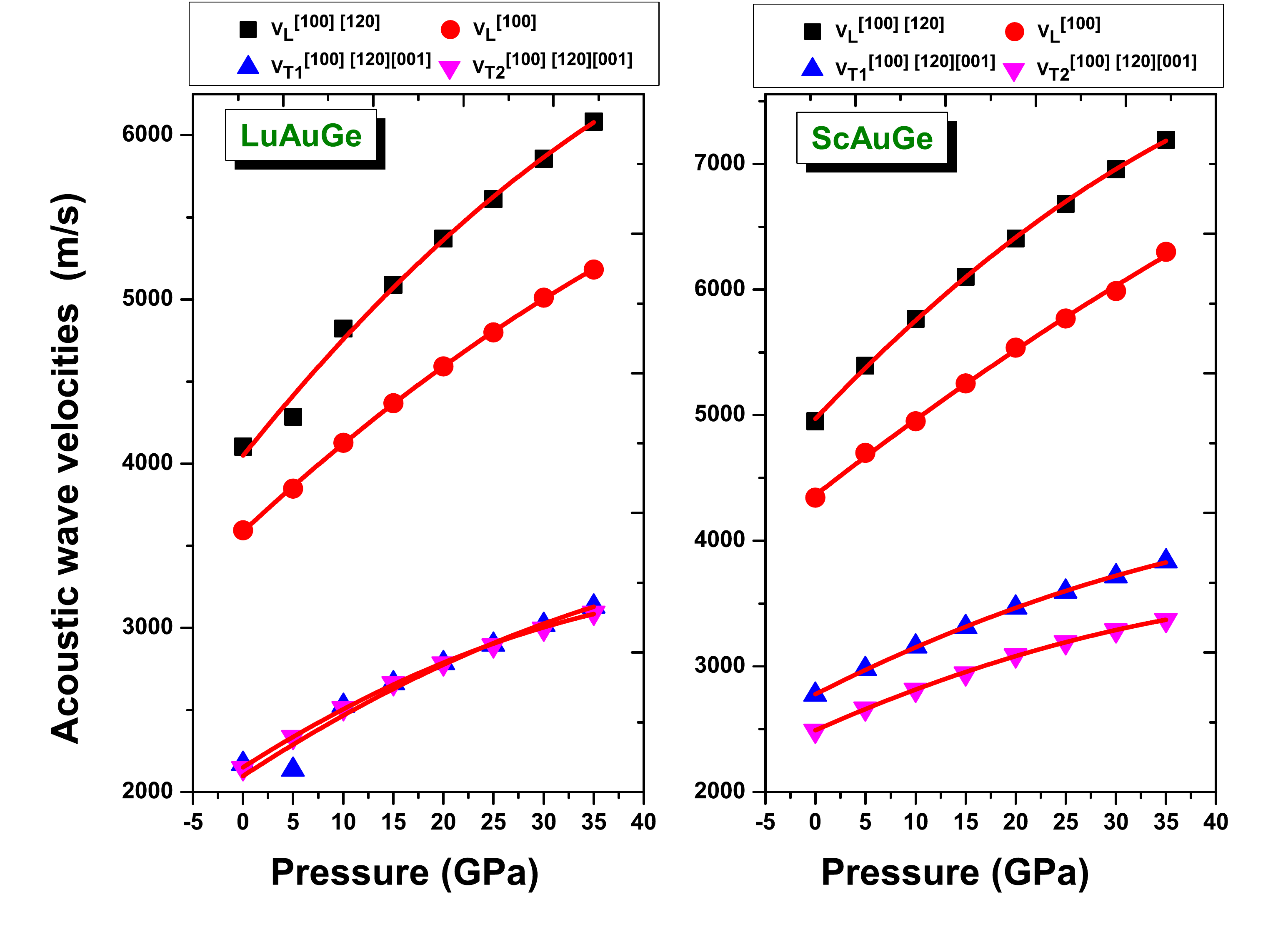}
    \end{center}
	\caption{(Colour online) Pressure dependence of acoustic wave velocities  for
different directions of propagation in the MAuGe (M = Lu, Sc) compounds.
The symbols indicate the calculated results. The lines represent the theoretical fits by a second-order polynomial.}
\label{fig7}
\end{figure}%

From  table~\ref{tab6}, we can see for both compounds that: (i) There
is a nuance between the values of the longitudinal velocities along the $a$%
-axis ($[100]$-direction) ($v_{L}^{[100]}=\sqrt{\frac{C_{11}}{\rho }}$) and $%
c$-axis ($[001]$-direction) ($v_{L}^{[001]}=\sqrt{\frac{C_{33}}{\rho }})$. The
longitudinal wave along the $a$-axis
travels faster than the longitudinal wave along the $c$-axis. (ii)
We can observe a difference between the values of the longitudinal
velocities along the $a$-axis ($v_{L}^{[100]}=\sqrt{%
\frac{C_{11}}{\rho }}$) and the transverse velocities along the $c$-axis ($v_{T_{1}}^{[001]}=v_{T_{2}}^{[001]}=\sqrt{\frac{C_{44}}{\rho }}$), the longitudinal wave along the $a$-axis travels faster
than the shear transverse wave because the square root of $C_{11}$ is larger
than $C_{44}$.

The sound wave velocities propagating in the $[100]$, $[001]$ and $[120]$
directions under the effect of pressure are shown in figure~\ref{fig7}.
This figure shows that all these acoustic wave velocities for different
propagations increase with an increase of the pressure and are well adjusted
by a second order polynomial equation for LuAuGe and
ScAuGe, respectively:

\begin{equation}
\text{LuAuGe}\left\{
\begin{array}{l}
V_{L}^{\left[ 001\right] }=3588.11+56.33P-0.03P^{2} \\
V_{T_{1}}^{\left[ 001\right] }=2150.96+38.87P-0.34P^{2} \\
V_{T_{2}}^{\left[ 001\right] }=2098.16+39.75P-0.29P^{2} \\
V_{L}^{\left[ 100\right] }=V_{L}^{\left[ 102\right]
}=4048.63+75.92P-0.51P^{2}%
\end{array}%
\right.
\end{equation}

\begin{equation}
\text{ScAuGe}\left\{
\begin{array}{l}
V_{L}^{\left[ 001\right] }=4363.16+62.24P-0.22P^{2} \\
V_{T_{1}}^{\left[ 001\right] }=2778.62+40.19P-0.29P^{2} \\
V_{T_{2}}^{\left[ 001\right] }=2490.99+35.36P-0.29P^{2} \\
V_{L}^{\left[ 100\right] }=V_{L}^{\left[ 102\right]
}=4970.27+84.17P-0.59P^{2}.%
\end{array}%
\right.
\end{equation}

\subsubsection{Elastic constants for polycrystalline aggregates}

The isotropic elastic parameters can fully describe  the mechanical behaviour of a polycrystalline material using one of the three pairs of isotropic elastic parameters: either the bulk modulus $B$ with the shear modulus $G$, the two Lam\'{e}'s constants $\lambda $ and $\mu$ or the Young's modulus $E$ with the Poisson's ratio $\nu $. Theoretically, the two isotropic elastic parameters $B$ and $G$ of the polycrystalline phase of a material can be
obtained by a special averaging of the individual elastic constants $C_{ij}s$ of the
monocrystalline phase. The Reuss--Voigt--Hill approximations \cite{Voigt28,Hill52} are the most used. Voigt $(B_{\text{V}}$, $G_{\text{V}})$ and Reuss $(B_{\text{R}}$, $G_{\text{R}})$ approximations represent the extreme values of $B$ and $G$ for polycrystalline samples. The two isotropic elastic parameters $B$ and $G$
are expressed for hexagonal systems as follows \cite{Wu07}:

\begin{equation}
\left\{
\begin{array}{l}
B_{\text{R}}=\frac{\left[ \left( C_{11}+C_{12}\right) C_{33}-2C_{13}^{2}\right] }{%
\left( C_{11}+C_{12}+2C_{13}-4C_{13}\right) } \\
B_{\text{V}}=\frac{2}{9}\left( C_{11}+C_{12}+C_{33}+\frac{C_{33}}{2}+2C_{13}\right)
\\
G_{\text{V}}=\frac{1}{30}\left( 7C_{11}-5C_{12}+12C_{44}+2C_{33}-4C_{13}\right) \\
G_{\text{R}}=\frac{5}{2}\left\{ \frac{\left[ \left( C_{11}+C_{12}\right)
C_{33}-2C_{13}^{2}\right] C_{44}C_{66}}{3B_{\text{V}}C_{44}C_{66}+\left[ \left(
C_{11}+C_{12}\right) C_{33}-2C_{13}^{2}\right] \left( C_{44}+C_{66}\right) }%
\right\}.%
\end{array}%
\right.  \label{12}
\end{equation}

Hill recommends that the arithmetic mean of the Voigt and Reuss limits
should be used in practice as an efficient model for determining the isotropic elastic parameters of polycrystalline samples:

\begin{equation}
\left\{
\begin{array}{c}
B_{\text{H}}=\frac{B_{\text{V}}+B_{\text{R}}}{2} \\
G_{\text{H}}=\frac{G_{\text{V}}+G_{\text{R}}}{2}\,,%
\end{array}%
\right.  \label{13}
\end{equation}
where $B_{\text{H}}$ and $G_{\text{H}}$ are the bulk and shear moduli, respectively, of the polycrystalline material according to Hill's approximation. The Young's modulus $E$ and the Poisson's ratio
$\nu $ for anisotropic material can be calculated from $B_{\text{H}}$ and $G_{\text{H}}$ using the following expressions:

\begin{equation}
\left\{
\begin{array}{c}
E=\frac{9B_{\text{H}}G_{\text{H}}}{3B_{\text{H}}+G_{\text{H}}} \\
\nu =\frac{3B_{\text{H}}-2G_{\text{H}}}{6B_{\text{H}}+2G_{\text{H}}}.%
\end{array}%
\right.   \label{14}
\end{equation}

The calculated bulk modulus $B_{\text{H}}$, shear modulus $G_{\text{H}}$, Young's
modulus $E$ and Poisson's ratio $\nu $ are listed in  table~\ref{fig7}.

\begin{table}[!t]%
\caption{Calculated bulk modulus ($B_\text{H}$, in GPa); shear modulus ($G_\text{H}$, in GPa); Young's modulus ($E$, in GPa) and Poisson's ratio ($\nu$) for the hexagonal compounds MAuLu (M = Lu, Sc) obtained using the single-crystal elastic constants $C_{ij}s$. The subscript R, V or H indicates that the modulus was obtained using the Reuss-Voigt-Hill approximations, respectively.}
\centering
\vspace{3mm}
\begin{tabular}[b]{|c|c|c|c|c|c|c|c|c|c|}
\hline
System & $B_\text{V}$ & $B_\text{R}$ & $B_\text{H}$ & $G_{\text{V}}$ & $G_\text{R}$ & $G_\text{H}$ & $B_\text{H}/G_{\text{H}}$ & $E$ & $\nu $ \\ \hline\hline
LuAuGe & 151.02 & 111.60 & 131.31 & 58.61 & 54.81 & 56.71 & 2.35 & 145.30
& 0.28 \\ \hline
ScAuGe & 157.28 & 114.22 & 135.75 & 67.29 & 62.26 & 64.78 & 2.11 & 168.20
& 0.26 \\ \hline
\end{tabular}%
\label{tab7}
\end{table}%

The reported results in table~\ref{tab7} allow us to make the following conclusions:
(i) From table~\ref{tab7}, it can be seen that the values of the bulk modulus deduced from the single-crystal elastic constants $C_{ij}s$ are in the same order of magnitude as their corresponding values calculated from the fit of the Pressure-Volume (P-V) data by different EOS  (see table~\ref{tab3}). (ii) Young's modulus E is used to provide a
measure of the stiffness of solids. Its values are 145.30~GPa (168.20 GPa) for LuAuGe (ScAuGe), which indicates the relatively noticeable resistance of MAuGe (M = Lu, Sc) to uniaxial deformation of compression/traction. (iii) The empirical Pugh criterion \cite{Pugh54} defined by the $B/G$ ratio is used to predict the ductile ($B/G > 1.75$) or brittle ($B/G < 1.75$) nature of materials. The value of the Pugh's ratios ($B_{\text{H}}/G_{\text{H}}$) using the Hill's approximation shown in table~\ref{tab7} for both LuAuGe and ScAuGe is greater than $1.75$, which suggests that both MAuGe are ductile. Thus, they will be resistant to thermal shocks. (iv) The Poisson's ratio $\nu $ is generally related to the volume change in a solid during uniaxial strain  \cite{Haddadi10,Ravindran98,Bouhemadou13}. From the values of $\nu $ in table~\ref{tab7}, the smallest calculated values are 0.28 for LuAuGe and 0.26 for ScAuGe, which shows that a considerable change in volume can be associated with elastic deformation in the considered materials.

For a complete description of the mechanical properties of MAuGe (
M = Lu, Sc), we also computed the isotropic longitudinal $V_{l}$, transverse $V_{t}$ and average $V_{m}$ sound wave velocities  using the following relations \cite{Anderson63,Schreiber74}:

\begin{equation}
\left\{
\begin{array}{c}
V_{m}=\left[ \frac{1}{3}\left( 2V_{t}^{-3}+V_{l}^{-3}\right) \right] ^{-1/3}
\\
V_{l}=\left( \frac{3B+4G}{3\rho }\right) ^{1/2} \\
V_{t}=\left( \frac{G}{\rho }\right) ^{1/2},
\end{array}%
\right.
\end{equation}
where, $B$ is the bulk modulus, $G$ is the shear modulus and $\rho $ is the
mass density. We also estimated the Debye temperature $\theta _\text{D}$, which is
an important physical parameter. The  Debye temperature $\theta _\text{D}$ is defined in terms of the average sound velocity $V_{m}$ as follows \cite{Anderson63,Ravindran98}:

\begin{equation}
\theta _\text{D}=\frac{h}{k_\text{B}}V_{m}\left[ \frac{3n}{4\piup }\frac{N_\text{A}\rho }{M}%
\right] ^{1/3},  \label{15}
\end{equation}
where, $h$ and $k_\text{B}$ are the Planck constant and Boltzmann constant, respectively, $N_\text{A}$ is the Avogadro number, $\rho $ is the mass density, $M$ is the molecular weight and $n$ is the number of atoms per unit cell.

The calculated sound velocities ($V_l$, $V_t$ and $V_m$) and Debye temperature $\theta_\text{D}$ values are reported in table~\ref{tab8}.

\begin{table}[!t]%
\caption{Calculated longitudinal, transverse and average sound
velocities ($V_l$, $V_t$ and $V_m$, in m/s), mass density  $\rho$ (g/cm$^3$) and Debye temperature ($\theta_\text{D}$, in K) for the hexagonal compounds MAuLu (M = Lu, Sc). }%
\centering
\vspace{3mm}
\begin{tabular}[b]{|c|c|c|c|c|c|}
\hline
System & $\rho $ & $V_{l}$ & $V_{t}$ & $V_{m}$ & $\theta _\text{D}$ \\
\hline\hline
LuAuGe & $12.51$ & $3876.95$ & $2133.22$ & $2377.63$ & $262.15$ \\ \hline
ScAuGe & $9.49$ & $4665.49$ & $2655.50$ & $2951.68$ & $333.09$ \\ \hline
\end{tabular}%
\label{tab8}
\end{table}%

\begin{figure}[!t] 
	\begin{center}
       \includegraphics[scale=0.50]{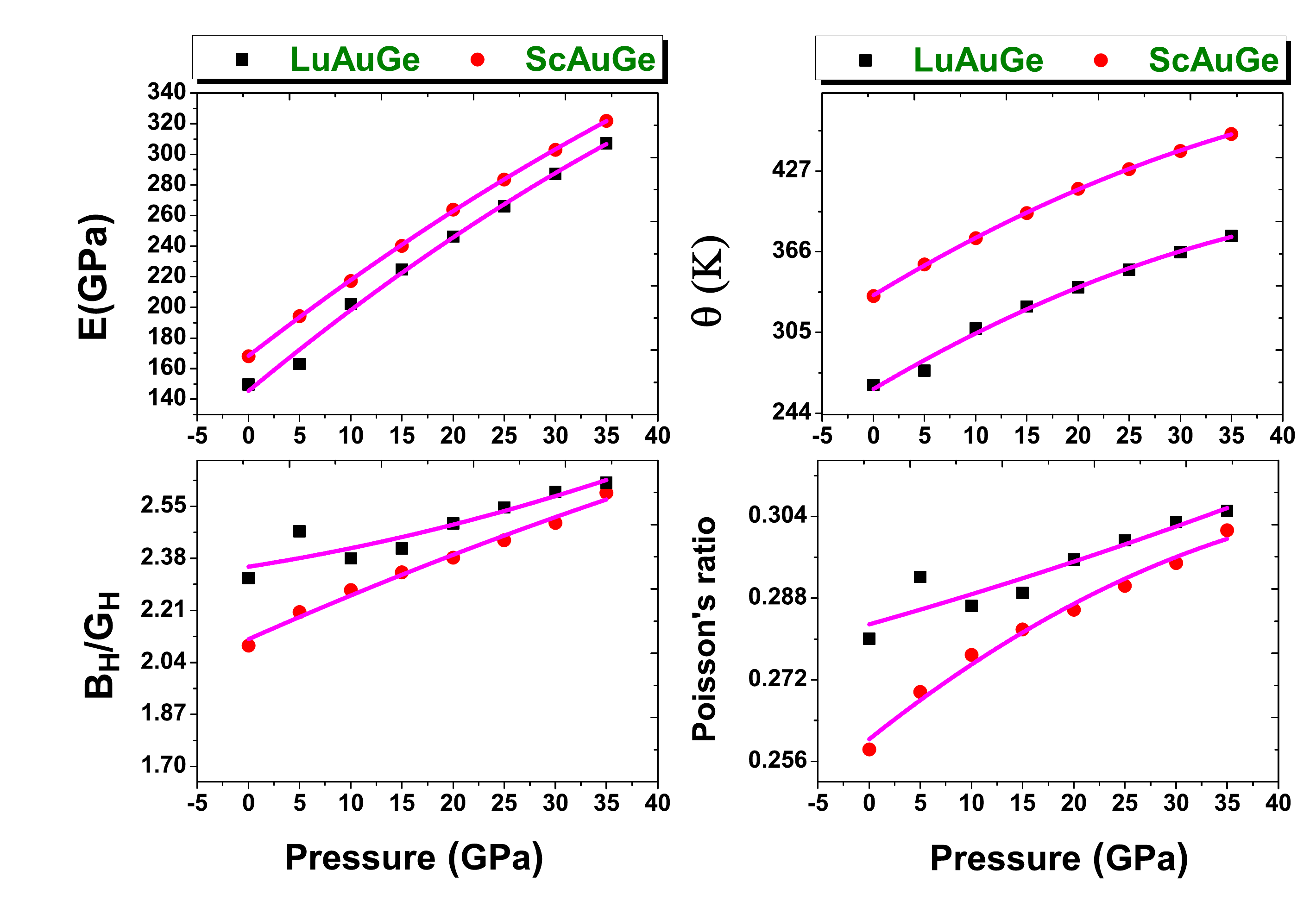}
    \end{center}
	\caption{(Colour online) Calculated pressure dependence of Pugh's ratio $B_\text{H}$/$G_\text{H}$, Young's
modulus E, Poisson's ratio $\nu$ and Debye temperature  $\theta_\text{D}$ for the hexagonal compounds MAuLu (M = Lu, Sc). The symbols indicate the calculated results. The lines represent the quadratic fit curves.}
\label{fig8}
\end{figure}

From table~\ref{tab8}, it is clear that the  Debye temperature and the speed of sound of LuAuGe are lower than those of  ScAuGe. The behavior of $B_{\text{H}}/G_{\text{H}}$, $E$, $\theta _\text{D}$, $\nu $, $V_{l}$, $V_{t}$ and $V_{m}$ under pressure is illustrated in  figure~\ref{fig8} and  figure~\ref{tab9}. From these two figures we can see a quadratic increase for all mentioned parameters with increasing the pressure. 

\begin{figure}[!t] 
\centering
      \includegraphics[scale=0.40]{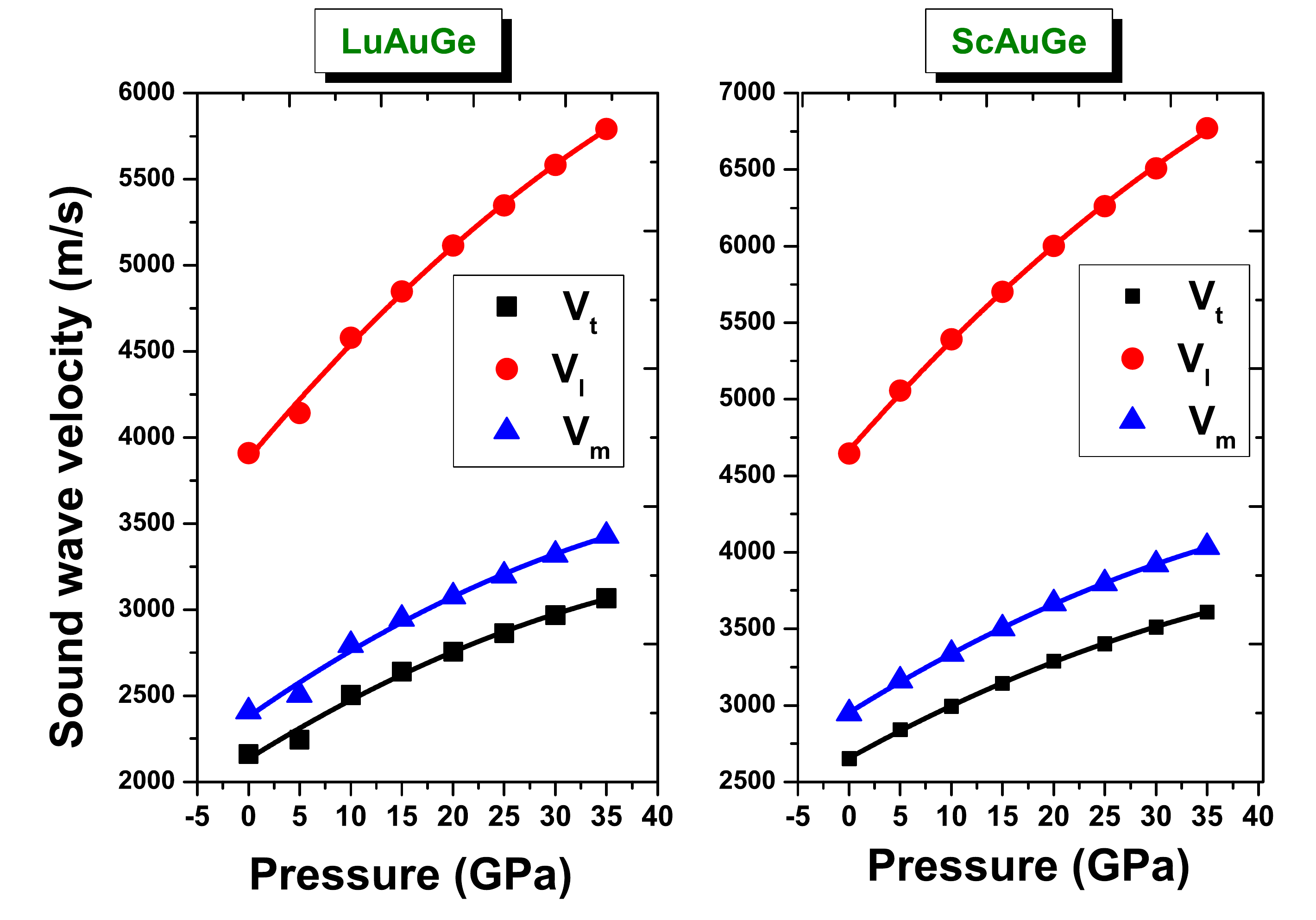}
	\caption{(Colour online) Calculated pressure dependence of  the sound (longitudinal $V_l$, transverse $V_t$ and average $V_m$)  for the hexagonal
compounds MAuLu (M = Lu, Sc). The symbols indicate the calculated results. The lines represent the quadratic fit curves.}
\label{fig9}
\end{figure}%

The resulting polynomial equations of $\ B_{\text{H}}/G_{\text{H}},$ $E,$ $\theta _\text{D},$ $\nu $, $V_{l}$, $V_{t}$ and $V_{m}$ for the MAuGe (M = Lu, Sc) compounds are listed as follows:

\begin{equation}
\text{LuAuGe}\left\{
\begin{array}{l}
\frac{B_{\text{H}}}{G_{\text{H}}}=2.35+5.24\times10^{-3}P-8.19\times10^{-5}P^{2} \\
E=145.30+5.55P-0.026P^{2} \\
\theta _\text{D}=262.15+4.58P-0.036P^{2} \\
\nu =0.282+5.65\times10^{-4}P-2.46\times10^{-6}P^{2} \\
V_{l}=3876.95+71P-0.46P^{2} \\
V_{t}=2133.22+37.13P-0.30P^{2} \\
V_{m}=2377.63+41.56P-0.33P^{2}
\end{array}%
\right.   \label{19}
\end{equation}

\begin{equation}
\text{ScAuGe}\left\{
\begin{array}{l}
\frac{B_{\text{H}}}{G_{\text{H}}}=2.11+1.47\times10^{-2}P-4.92\times10^{-5}P^{2} \\
E=168.20+5.19P-0.023P^{2} \\
\theta _\text{D}=333.09+4.96P-0.034P^{2} \\
\nu =0.26+1.6\times10^{-3}P-1.36\times10^{-5}P^{2} \\
V_{l}=4665.49+76.28P-0.47P^{2} \\
V_{t}=2655.50+36.86P-0.27P^{2} \\
V_{m}=2951.68+41.59P-0.30P^{2}.%
\end{array}%
\right.
\end{equation}

\subsubsection{Elastic anisotropy}

Elastic anisotropiy has an important implication in the engineering science. Recent research shows that the elastic anisotropy for solid crystals has an influence on 
microcracks in materials \cite{Ravindran98,Chung68} and on the nanoscale
precursor textures in alloys \cite{Lloveras08,Rong13}. Different criteria
have been developed to describe the elastic anisotropy of materials.
(i) For a hexagonal structure, the anisotropic shear factors $A_{1}$, $A_{2}$, and $%
A_{3}$ provide a measure of the degree of anisotropy for the bonding
between atoms in different planes. For an isotropic crystal, $A_{1}$, $A_{2}$ and $A_{3}$ should have values equal to unity, while any value other than unity is an indication of elastic anisotropy. The elastic anisotropy factors $A_{1}$, $A_{2}$ and $A_{3}$ can be expressed as follows:

\begin{equation}
\left\{
\begin{array}{l}
A_{1}=A_{2}=4C_{44}/\left( C_{11}+C_{33}-2C_{13}\right) \text{ \ \ \ \ \ \
For the }\left( 100\right) \text{ and }(010)\text{ planes,} \\
A_{3}=4C_{66}/\left( C_{11}+C_{22}-2C_{12}\right) \text{ \ \ \ \ \ \ \ \ \ \
\ \ \ \ \ For the }(001)\text{ plane.} \\
C_{66}=(C_{11}-C_{12})/2%
\end{array}%
\right.
\end{equation}

(ii) Another way to evaluate the elastic anisotropy consists in introducing the
Voigt and Reuss bounds~\cite{Chung68}. The elastic anisotropy in compression (shear) defined by the factor $A_{B}$ ($A_{G}$) is expressed as follows:

\begin{equation}
\left\{
\begin{array}{c}
A_{B}($\%$)=\frac{B_{\text{V}}-B_{\text{R}}}{B_{\text{V}}+B_{\text{R}}}\times 100 \\
A_{G}$(\%$)=\frac{G_{\text{V}}-G_{\text{R}}}{G_{\text{V}}+G_{\text{R}}}\times 100,%
\end{array}%
\right.
\end{equation}
where $B$ and $G$ are the bulk and shear moduli, and the subscripts V and R represent the Voigt and Reuss bounds. The $A_{B}$ and $A_{G}$ ratios can
range from zero to $100\%$. A value of zero represents elastic isotropy and
a value of $100\%$ represents the largest possible elastic anisotropy. 

(iii) The third way consists in precise quantifying the extent of the elastic anisotropy using the universal index $A^{U}$ \cite{Ranganathan08}. The index $A^{U}$ takes into  account both compression and shear contributions, which is defined as follows:

\begin{equation}
A^{U}=5\frac{G_{\text{V}}}{G_{\text{R}}}+\frac{B_{\text{V}}}{B_{\text{R}}}-6.
\end{equation}

The universal index is equal to zero $(A^{U}=0)$ for isotropic crystals, and
the deviation of $A^{U}$ from zero shows the presence of elastic
anisotropy.

The elastic anisotropy values deduced from the factors $A_{1}$, $A_{2}$, $A_{3}$, $A_{B}$, $A_{G}$ and $A^{U}$ are given in  table~\ref{tab9}. From this table, we see that $A_{3}$ values indicate an isotropy in the shear plane $(001)$  while the values from $A_{1}$ and $A_{2}$ indicate the presence of a very low anisotropy in the shear plane $(100)$ and $(010)$ for the MAuGe (M = Lu, Sc) compounds. The $A_{B}$ and $A_{G}$ values show that the elastic compression anisotropy is relatively more pronounced than the  shear anisotropy for both compounds. The $A^{U}$ values also indicate the presence of elastic anisotropy for the two studied materials.
\begin{figure}[!t] 
\centering%
      \includegraphics[scale=0.65]{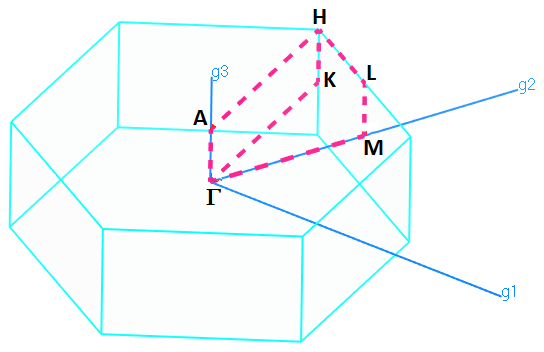}
	\caption{(Colour online) The first Brillouin zone of MAuGe (M = Lu, Sc). The dashed red lines are the selected path for band structure calculation. $g_1$, $g_2$ and $g_3$ are the reciprocal lattice axes.}
	\label{fig10}
\end{figure}%

\begin{table}[!t]%
\caption{Calculated elastic anisotropy factors : A$_1$, A$_2$, A$_3$, A$_B$, A$_G$, and A$^U$ for MAuGe (M = Lu, Sc).} 
\centering
\vspace{3mm}
\begin{tabular}[b]{|c|c|c|c|c|c|c|}
\hline
System  & $A_{1}$ & $A_{2}$ & $A_{3}$ & $A_{B}$ ($\%$) & $A_{G}$ ($\%$)& $A^{U}$ \\ \hline
LuAuGe & 0.96 & 0.96 & 1 & 15 & 3.3 & 0.69  \\ \hline
ScAuGe & 0.80 & 0.80 & 1 & 15.85 & 3.38 & 0.78  \\ \hline
\end{tabular}%
\label{tab9}
\end{table}%

\subsection{Electronic properties}

\begin{figure}[!t] 
\centering%
       \includegraphics[scale=0.51]{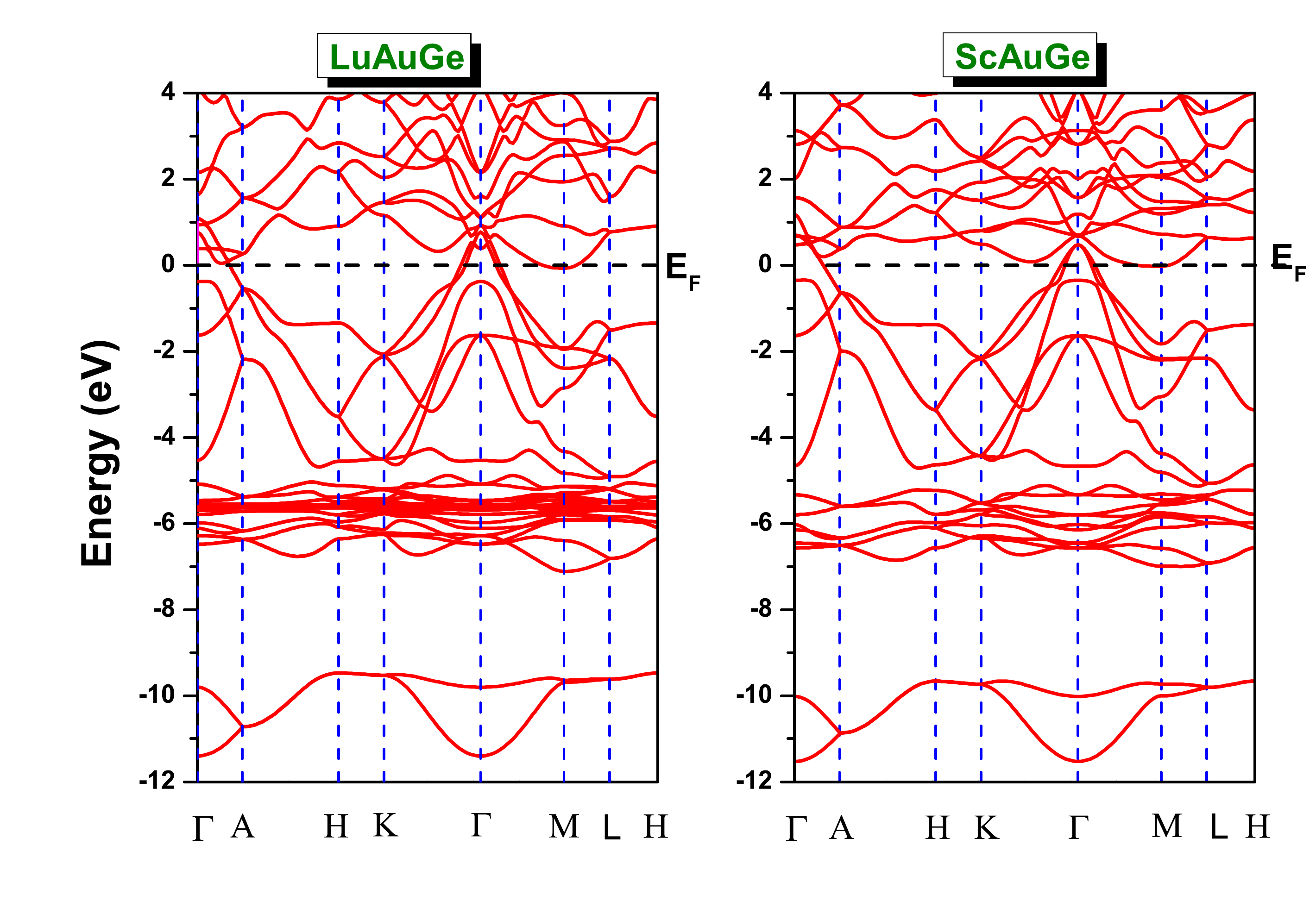}
	\caption{(Colour online) Electronic band dispersion curves along the high symmetry directions in the Brillouin zone for the MAuGe (M = Lu, Sc) compounds.}
	\label{fig11}
\end{figure}

The Brillouin zone (BZ) which highlights the selected path $\Gamma$-A-H-K-$\Gamma$-M-L-H to calculate the energy band structures for the MAuGe (M=Lu, Sc) compounds, is illustrated in figure~\ref{fig10}. The MAuGe band structures along  the chosen path are illustrated in figure~\ref{fig11}. It is seen that the LuAuGe and ScAuGe compounds at their equilibrium lattice parameters have similar energy band dispersions in the considered energy range (from $-12$~eV to 4~eV) with some small differences depending on the electron valence states of the Lu/Sc,Au and Ge atoms. It should be noted that the valence and conduction bands overlap at the Fermi level (EF), which reveals the absence of the bandgap at the Fermi level. Consequently, the MAuGe (M = Lu, Sc) compounds exhibit a metallic nature.

\begin{figure}[!t] 
\centering
      \includegraphics[scale=0.50]{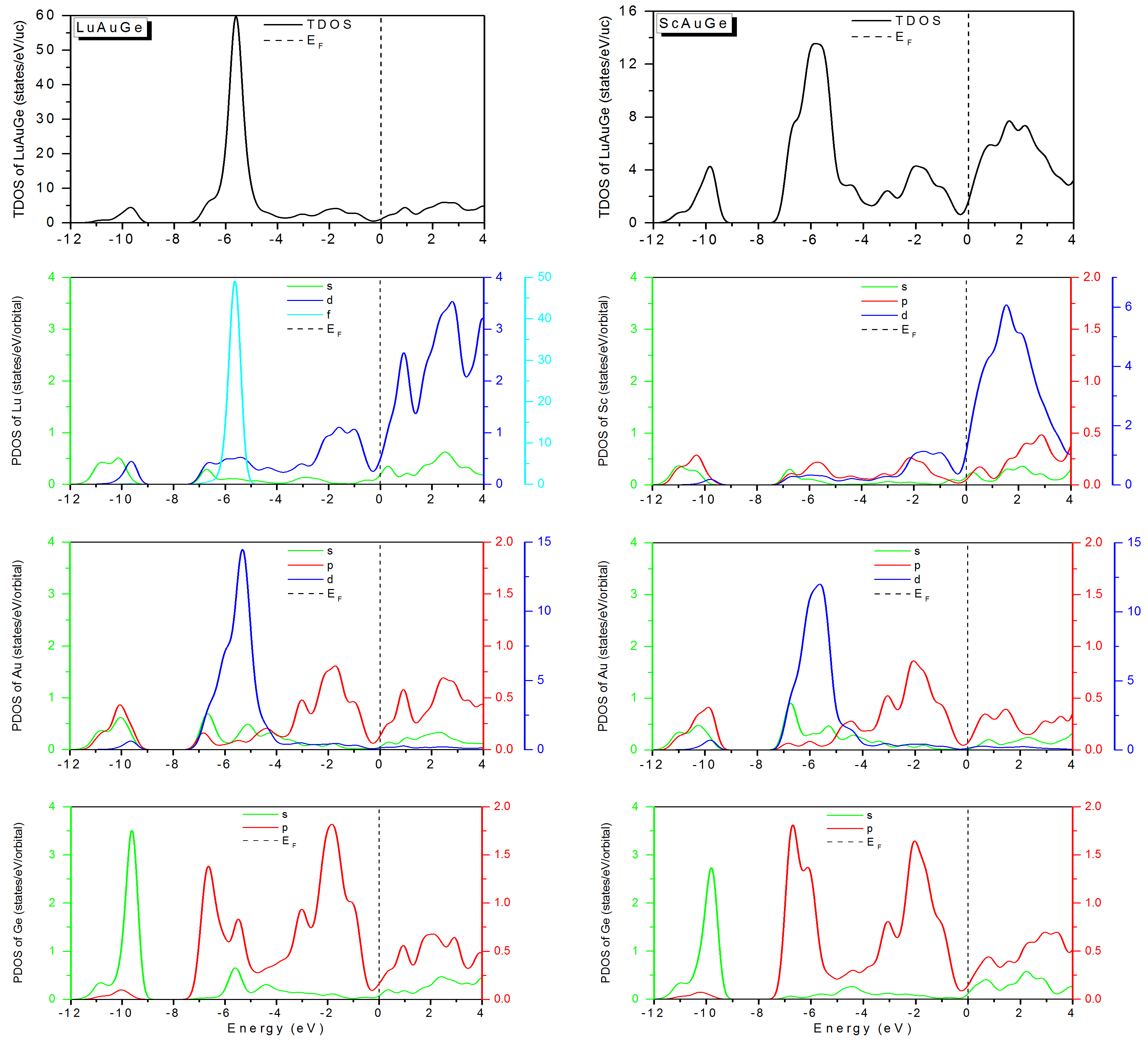}
	\caption{(Colour online) Total (TDOS) and partial (PDOS) densities of states diagrams for MAuGe (M = Lu, Sc).}
		\label{fig12}
\end{figure}

In order to determine the contribution of the electron valence states of each atom in the MAuGe electronic band structures, we calculated the total (TDOS) and partial (PDOS) densities of states for both compounds. The TDOS and PDOS diagrams are shown in figure~\ref{fig12}. It is clear that the total density of states for MAuGe is characterized by two distinct regions in the $-12$~eV to 0~eV energy range. The first region is located between $-12$ and $-9$ eV and is mostly derived from the Ge-s states for both studied compounds. The second region starts from about $-7.5$ eV up to the Fermi level ($E_F$) and is mainly composed of the Lu-f and Au-d and Ge-p states.  The lowest conduction bands (from 0 to 4~eV) are mostly made up of the Lu/Sc-d unoccupied states.

\begin{figure}[!t] 
\centering%
	\begin{center}
      \includegraphics[scale=0.45]{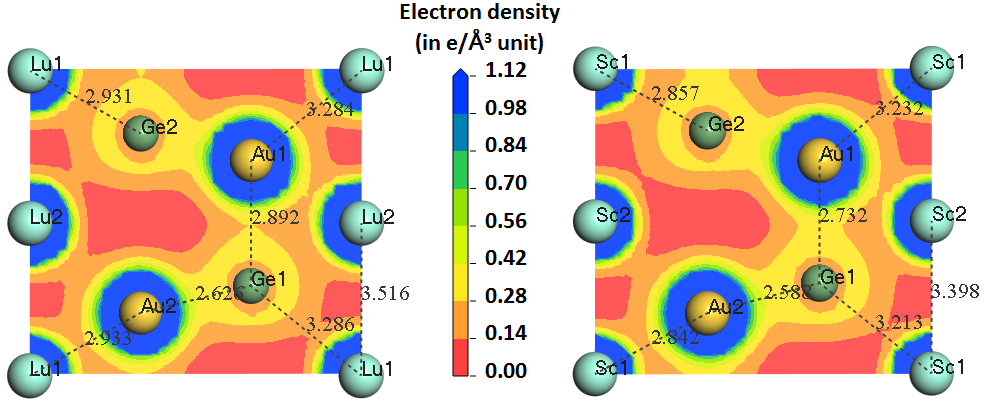}
    \end{center}
	\caption{(Colour online) Electron density  maps in the (110) crystallographic plane for the  MAuGe (M = Lu, Sc) compounds. The electron density is high in the blue regions (+1.12 e/\AA$^3$) and it is low in the red regions (+0.00 e/\AA$^3$). The color density scale is in electrons/\AA$^3$.}
		\label{fig13}
\end{figure}

To better understand the chemical bonding character between the Lu/Sc, Au and Ge atoms, the electron density distribution maps in the crystallographic plane (110) for MAuGe (M = Lu, Sc) compounds are shown in figure~{\ref{fig13}} together with the calculated bond lengths that are listed in table~\ref{tab4}. One can see that the Ge1-Au2 and Ge1-Au1 bonds are characterized by an obvious deformation of the electron charge density distribution (see the yellow area in the electron density maps between Ge and Au atoms), which indicates that the Ge1-Au2 and Ge1-Au1 bonds have a covalent bonding nature for both compounds studied. The hybridization of the Ge-p and Au-p states, which is clearly visible in the PDOS spectra shown in figure 12, is responsible for the Ge-Au covalent bonds. It is worth to note that the covalent bonding between Ge1 and Au1 is more pronounced in the ScAuGe compound. The electron charge density is typically low and uniform (see the electron density area whose values are between 0.14 and 0.28 e/\AA$^3$) along the (Lu1/Sc1)-Au2, (Lu1/Sc1)-Au1, (Lu1/Sc1)-Ge2, (Lu1/Sc1)-Ge1 and Lu1-Lu2 (Sc1-Sc2), which indicates the presence of the metallic character between these bonds.
 The metallic bonding is attributed to the presence of the delocalized Lu/Sc-d states.

\begin{table}[!t]%
\caption{Hirshfeld's atomic charges of Lu/Sc, Au and Ge atoms in the MAuGe (M = Lu, Sc) compounds.}
\centering
\vspace{3mm}
\begin{tabular}[b]{|c|c|c|c|c|c|c|c|}
\hline
& Atom& Lu1 & Lu2 & Ge1 & Ge2 & Au1 & Au2 \\ \cline{2-8}
LuAuGe &Charge & 0.07 & 0.07 & $-0.14$ & $-0.14$ & 0.08 & 0.08\\ \hline
& Atom& Sc1 & Sc2 & Ge1 & Ge2 & Au1 & Au2  \\ \cline{2-8}
ScAuGe &Charge & 0.01 & 0.01 & $-0.12$ & $-0.12$ & 0.11 & 0.11 \\  \hline
\end{tabular}%
\label{tab10}
\end{table}%

To further explore the MAuGe (M = Lu, Sc) electronic structures, we calculated the atomic charges of M, Au and Ge atoms using Hirshfeld's population analysis \cite{Hirshfeld77}. The obtained results are tabulated in  table~\ref{tab10}. One can observe that M = Lu/Sc, Au and Ge atoms have small effective charges (positive charges for M and Au, and negative charge for Ge). A very lower atomic charge difference implies much lower ionicity and higher covalency in the corresponding chemical bonds. Thus, the chemical bonding between Ge and Au is covalent.
 \newpage

\section{Conclusions}
In summary, we have performed ab initio calculations of the structural,  electronic and elastic properties for the MAuGe (M = Lu, Sc) compounds by means of the pseudopotential plane-wave method in the framework of the density functional theory within the generalized gradient approximation. Our results can be summarized as follows:
\begin{itemize}
\item[$\triangleright $] The optimized structural parameters are in very good agreement with the existing experimental and theoretical data.

\item[$\triangleright $] The elastic constants of the monocrystalline phase calculated at zero pressure show that the MAuGe (M = Lu, Sc) materials are mechanically stable. Note that the mechanical stability remains verified for hydrostatic pressures up to 40~GPa.

\item[$\triangleright $] The numerical estimates of the elastic moduli of the polycrystalline phase, i.e., Young's modulus, shear modulus, Poisson's ratio, anisotropy factors, sound velocities and Debye temperature were evaluated and discussed under pressure for the first time. 
%The Pugh's ratio that LuAuGe and ScAuGe are ductile materials.
The Pugh's ratio for LuAuGe and ScAuGe indicates that these materials are ductile.

\item[$\triangleright $] The electronic structures analysis shows that the MAuGe (M = Lu, Sc) compounds are of a metallic character. This behavior is attributed to the delocalized d states of the Lu and Sc atoms. According to the densities of states and the electron charge maps in the (110) plane, it has been deduced that there are covalent interactions between the Au and Ge atoms.
\end{itemize}

\section*{Acknowledgements}
The authors express their thanks to Drs H. Zitouni, M. Ahmed Ammar, N. Zaghou, T. Bitam and D.~Houatis for their help, support, constant assistance and for their advice throughout the realization of this paper.

\newpage
\ukrainianpart
\title{Ab initio вивчення структурних, пружних та електронних властивостей гексагональних MAuGe (M= Lu, Sc) сполук}
%\thanks{Missoum RADJAI: E-mail: mradjai@yahoo.com}}
%
\author{М. Раджаї\refaddr{label1},
      Н. Гуечі\refaddr{label2,label3}, Д. Моуче \refaddr{label4}}
\addresses{
\addr{label1} Лабораторія фізики експериментальної техніки і їх застосування (LPTEAM), університет Медеа, Алжир
\addr{label2} Лабораторія досліджень поверхонь твердих матеріалів та інтерфейсів, університет Ферхат Аббас Сетіф 1, Алжир 
\addr{label3} Медичний факультет, університет Ферхат Аббас Сетіф 1, Алжир
\addr{label4} Лабораторія  нових матеріалів та їх характеристик, університет Ферхат Аббас Сетіф 1, Алжир
}

\makeukrtitle 
\begin{abstract}
У статті проведено детальне теоретичне дослідження структурних, пружних та електронних властивостей двох германідів LuAuGe та ScAuGe за допомогою першопринципних розрахунків  із використанням методу псевдопотенціальної плоскої хвилі в рамках  узагальненого градієнтного наближення. Параметри кристалічної гратки та внутрішні координати добре узгоджуються з існуючими експериментальними та теоретичними даними, що підтверджує надійність застосовуваного теоретичного методу. Показано вплив гідростатичного тиску на  структурні параметри. Монокристалічні пружні константи розраховували за допомогою деформації напругової техніки. Розраховані пружні константи сполук MAuGe (M = Lu, Sc) відповідають критеріям механічної стійкості для гексагональних кристалів, і ці константи використовувались для аналізу пружної анізотропії сполук MAuGe за трьома різними показниками. Полікристалічні ізотропні модулі пружності, а саме об'ємний модуль, модуль зсуву, модуль Юнга, коефіцієнт Пуассона та відповідні властивості також оцінюються за допомогою наближень Фойгта-Ройсса-Хілла. Також ми дослідили електронні властивості розглянутих сполук шляхом обчислення їх зонних структур, їх щільності станів та розподілу електронної густини.

\keywords LuAuGe, ScAuGe, метод PP-PW, електронні властивості, модулі пружності, ab initio розрахунки 
\end{abstract}%

\lastpage

\end{document}